# Molecular Environment of the Thermal Composite Supernova Remnant G352.7-0.1

Qian-Qian Zhang[1], Ping Zhou[1,2], Yang Chen[1,2], Xiao Zhang[1,2], Wen-Juan Zhong[1], Xin Zhou[3], Zhi-Yu Zhang[1,2], and Jacco Vink[4]
[1] School of Astronomy & Space Science, Nanjing University, 163 Xianlin Avenue, Nanjing 210023, People's Republic of China; pingzhou@nju.edu.cn ygchen@nju.edu.cn and xiaozhang@nju.edu.cn
[2] Key Laboratory of Modern Astronomy and Astrophysics, Nanjing University, Ministry of Education, Nanjing 210023, People's Republic of China
[3] Purple Mountain Observatory and Key Laboratory of Radio Astronomy, Chinese Academy of Sciences, 10 Yuanhua Road, Nanjing 210023, People's Republic of China
[4] Anton Pannekoek Institute for Astronomy & GRAPPA, University of Amsterdam, Science Park 904, 1098 XH Amsterdam, The Netherlands

## Abstract

Galactic supernova remnants (SNRs) play an important role in our understanding of supernovae and their feedback on the interstellar environment. SNR G352.7-0.1 is special for its thermal composite morphology and double-ring structure. We have performed spectroscopic mapping of the $^{12}$CO and $^{13}$CO $J = 2$–1 lines toward G352.7-0.1 with the Atacama Pathfinder Experiment telescope. Broad $^{12}$CO lines are found in the northeastern ring at a local-standard-of-rest velocity range of $\sim -50$ to $-30$ km s$^{-1}$, suggesting that the remnant is interacting with molecular clouds at $\sim -51$ km s$^{-1}$. Thus, we adopt a distance of $\sim 10.5$ kpc for this SNR. The momentum and kinetic energy of the shocked gas along the line of sight are estimated to be $\sim 10^2 M_\odot$ km s$^{-1}$ and $\sim 10^{46}$ erg, respectively. We also find an expanding structure around the remnant, which is possibly related to the wind-blown bubble of the progenitor star. From Fermi-LAT data in the energy range 0.1–500 GeV, we find no gamma-ray counterparts to G352.7-0.1.

*Unified Astronomy Thesaurus concepts:* Interstellar medium (847); Supernova remnants (1667); Molecular clouds (1072); Shocks (2086)

## 1. Introduction

Molecular clouds (MCs) are the dense and cold phase of the interstellar medium (ISM), with a typical density of $10^2$–$10^6$ cm$^{-3}$ (Draine 2011). Massive stars, which are the progenitors of core-collapse supernovae (SNe), are usually not far from their parental MCs due to their short lifetimes, so core-collapse supernova remnants (SNRs) are expected to interact with ambient MCs. The interplay between SNRs and the molecular environment is an important topic. As the SN shock propagates in MCs, it alters the physical conditions and chemical compositions of the molecular gas. Generations of SNe are strong mechanical energy sources and strongly regulate the star formation rate (Keller & Kruijssen 2022). Moreover, a subgroup of SNRs, thermal composite SNRs (also known as mixed-morphology supernova remnants (MMSNRs)), is found to show a good correlation with MCs (Green et al. 1997; Zhang et al. 2015). MMSNRs are defined for those SNRs with a shell-like morphology in the radio band and centrally filled morphology in the X-ray band (Jones et al. 1998; Rho & Petre 1998). So far, most MMSNRs are core-collapse SNRs (Zhang et al. 2015).

The interaction with dense medium also strongly shapes the SNR morphology. Some core-collapse SNRs show highly asymmetric morphology in multiple wave bands. In contrast, Type Ia SNRs appear to be more symmetric and are supposed to evolve in a homogeneous medium (Lopez et al. 2011). The evolution delay time between star formation and Type Ia SN explosion is typically 100 Myr to several gigayears, in which time the system is expected to have escaped its initial environment, or the circumstellar medium has been altered by the winds and explosions of massive stars. Therefore, a nonmolecular environment for Type Ia SNe is expected. However, in recent years, molecular environment studies of Type Ia SNRs have revealed that they can also evolve in a dense medium, such as Tycho (Zhou et al. 2016) and N103B (Sano et al. 2018).

SNR G352.7−0.1 is an MMSNR with an asymmetric morphology (Giacani et al. 2009). It was first identified by Clark et al. (1973) according to observations at 408 MHz and 5000 MHz. Caswell et al. (1983) mapped the remnant with the Fleurs Synthesis Telescope and found its shell structure. Very Large Array (VLA) observations at 1.4 GHz revealed that the shell is actually composed of two overlapping rings (Dubner et al. 1993), and this double-ring structure was later confirmed by VLA 4.8 GHz observations with higher angular resolution (Giacani et al. 2009).

X-ray observations toward G352.7−0.1 have been conducted with the Advanced Satellite for Cosmology and Astrophysics, XMM-Newton, Chandra, and Suzaku since 1998 (Kinugasa et al. 1998; Giacani et al. 2009; Pannuti et al. 2014; Sezer & Gök 2014). The centrally filled thermal X-ray emission with a radio shell suggests that it is an MMSNR (Giacani et al. 2009; Pannuti et al. 2014). The progenitor of G352.7−0.1 is still under debate. Giacani et al. (2009) and Pannuti et al. (2014) favor a massive progenitor due to the asymmetric morphology and high swept-up mass, while the Fe-rich ejecta in a low ionization state supports a Type Ia SN explosion (Sezer & Gök 2014).

Since many MMSNRs are associated with MCs (Green et al. 1997), it is of interest to investigate whether there is SNR–MC interaction in G352.7−0.1, which yet remains uncertain. One of the most robust tracers of SNR–MC interaction is the 1720 MHz OH maser. Green et al. (1997) reported a survey of 75 Galactic SNRs with the Parkes 64 m telescope, where 1720 MHz OH line emission toward G352.7−0.1 was detected at a local-standard-of-rest (LSR) velocity $V_{LSR} \sim 0$ km s$^{-1}$. Further investigation tried to distinguish the compact shock-excited maser from extended Galactic thermal OH emission, and the result was not positive for this SNR (Koralesky et al. 1998). Pannuti et al. (2014) presented a Spitzer 24 $\mu$m image of the remnant and found that its infrared morphology agrees well with its radio morphology, which indicates that the SNR is evolving in an inhomogeneous medium.

Motivated by the intriguing double-ring, thermal composite morphology of SNR G352.7−0.1, we have performed molecular observations toward the remnant. Our goal is not only to study the origin of the SNR morphology but also to investigate the SNR properties and how much SN energy is transferred to the ISM. Since SNR–MC interactions are also thought to be responsible for hadronic gamma rays, we also explore Fermi-LAT data to determine whether or not the remnant has a gamma-ray counterpart.

In Section 2, we introduce our observations and data. In Section 3, we present our main results, and the results are discussed in Section 4. A summary is given in Section 5.

## 2. Observations and Data Reduction

### 2.1. APEX Observation

While H$_2$ molecules do not have a permanent electric dipole moment, CO molecules are the most practical tracer of molecular gas and the most common probe of the physical properties of MCs. The observations in $^{12}$CO $J=2$–1 (230.538 GHz) and $^{13}$CO $J=2$–1 (220.399 GHz) toward G352.7−0.1 were carried out with the Atacama Pathfinder Experiment (APEX) on 2019 May 7 (Program ID: 0103.D-0387(A); PI: P. Zhou). Compared to CO $J=1$–0, CO $J=2$–1 has a shorter wavelength, resulting in a higher spatial resolution. In addition, the emission of CO $J=2$–1 is expected to be stronger than that of $J=1$–0 if there is shock heating (Seta et al. 1998). We mapped an area of about $12' \times 12'$ centered at ($17^h27^m40^s$, $-35°07'40''$, J2000) using the on-the-fly mode. The pointing accuracy was better than $2''$. We used $\alpha(2000) = 17^h38^m47^s$, $\delta(2000) = -35°27'13''$ as the reference point. The data were obtained by nFLASH receiver with a main-beam efficiency of 0.77. The spacing of the scans was $9''$. The angular resolution was $27''$ at 230 GHz, and the frequency resolution was 61 kHz ($\sim 0.08$ km s$^{-1}$ at 230 GHz).

We use the GILDAS/CLASS package developed by IRAM[5] to reduce the data, producing cubes with a velocity resolution of 0.5 km s$^{-1}$ and grid spacing of $10''$. The mean rms noise (defined in $T_{mb}$) is about 0.1 K for both $^{12}$CO $J=2$–1 and $^{13}$CO $J=2$–1.

### 2.2. Fermi-LAT Data

We use the Fermipy (v1.1.2) package (Wood et al. 2017) to analyze about 14 yr (from 2008 August 4 15:43:36 UTC to 2022 June 1 05:12:31 UTC) of Fermi-LAT Pass 8 data to search for gamma-ray emission of SNR G352.7−0.1, applying the instrument response functions (IRFs) P8R3_SOURCE_V3_v1. The region of interest (ROI) is centered at the SNR (R.A. = 261°.92, decl. = −35°.112) with a size of $10° \times 10°$. The background model includes sources in 4FGL-

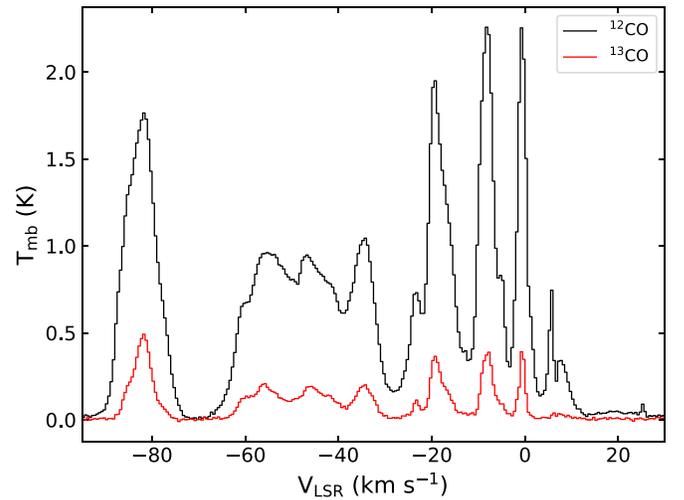

**Figure 1.** The averaged $^{12}$CO $J=2$–1 and $^{13}$CO $J=2$–1 spectra of G352.7−0.1 in the velocity range from $-95$ to $+30$ km s$^{-1}$.

DR3 within 15° from the ROI center, Galactic diffuse background emission (gll_iem_v07.fits), and isotropic emission (iso_P8R3_SOURCE_V3_v1.txt).

### 2.3. Other Wave Bands

To compare the molecular emission with multiwavelength images, we use the VLA 4.8 GHz radio continuum emission data provided by Giacani et al. (2009).

We also use H I and 1.4 GHz continuum data from the Southern Galactic Plane Survey (SGPS) to determine the distance to the SNR/MCs. The observations of H I and continuum emission were both made by the Australia Telescope Compact Array (ATCA) and Parkes Radio Telescope (McClure-Griffiths et al. 2005; Haverkorn et al. 2006). The velocity resolution of the spectra is 0.82 km s$^{-1}$.

## 3. Results

### 3.1. Distribution of the MCs

Figure 1 shows the $^{12}$CO and $^{13}$CO spectra averaged in a region of about $0°.2 \times 0°.2$ covering the remnant. Peaks at $V_{LSR} \sim -82, -55.5, -47, -34.5, -23.5, -19.5, -8.5, -1, +5.5$, and $+7.5$ km s$^{-1}$ are seen in the $^{12}$CO emission. The line profile of $^{13}$CO is similar to that of $^{12}$CO, though the peaks at $+5.5$ and $+7.5$ km s$^{-1}$ are not prominent.

We present integrated intensity maps of $^{12}$CO at different LSR velocities in Figure 2. To describe the distribution of molecular gas better, we divide the radio morphology of G352.7−0.1 into three parts: an eastern shell, a middle shell, and a western shell (as labeled in the first panel of Figure 2). The eastern and middle shells constitute the northeastern ring (also referred to as the inner shell/ring in previous studies), while the western shell (also referred to as the outer shell/arc) is part of the southwestern ring. There are two ledge-like structures (at $\delta \sim -35°06'$) on the eastern and middle shells, and we regard them as part of the southwestern ring.

The spatial distribution of the $^{12}$CO emission varies greatly with the velocity. In the LSR velocity interval of $-85$ to $-80$ km s$^{-1}$, the $^{12}$CO emission mainly lies to the west of G352.7−0.1 and overlaps with the southwestern region of the remnant. In the $-50$ to $-40$ km s$^{-1}$ interval, there are molecular patches in spatial correspondence with the middle shell. A linear structure can also be seen across the western and middle shells at $V_{LSR} = -45$ to $-40$ km s$^{-1}$. In the velocity interval of $-40$ to $-35$ km s$^{-1}$, a cavity-like structure overlaps with the remnant in the north and east, though not corresponding well with the radio shell. At $V_{LSR} = -25$ km s$^{-1}$, the CO emission shows spatial correlation with the eastern shell. At $V_{LSR} \geqslant -20$ km s$^{-1}$, diffuse emission is distributed all over the field of view (FOV), with complex molecular structures overlapping with the remnant in different regions. We argue that only the clouds at $V_{LSR} = -50$ to $-40$ km s$^{-1}$ interact with the remnant directly (see Section 3.2).

---
[5] http://www.iram.fr/IRAMFR/GILDAS/

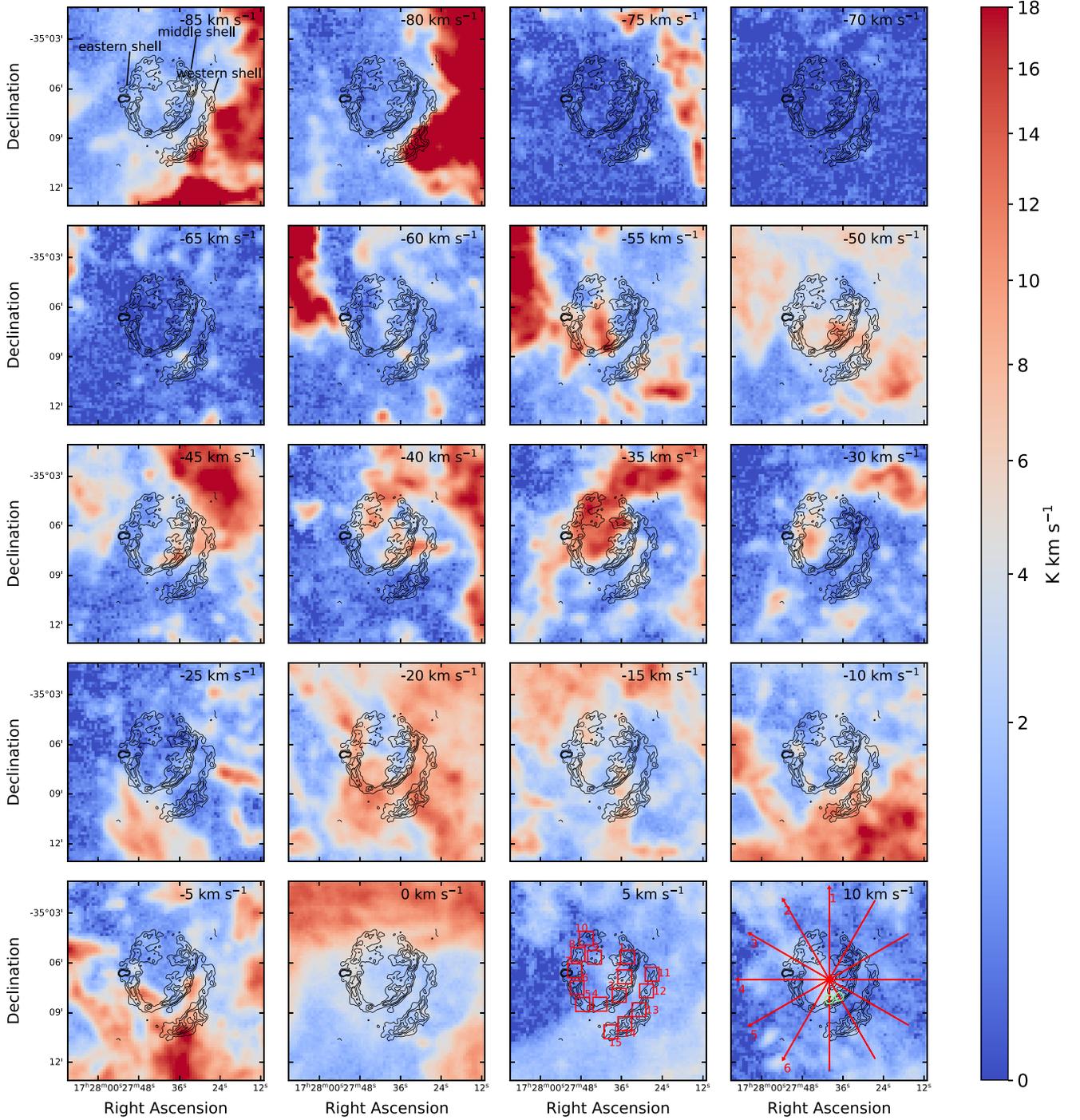

**Figure 2.** The intensity maps of $^{12}$CO $J = 2$–1 integrated every 5 km s$^{-1}$ from $-85$ to $+10$ km s$^{-1}$, overlaid with contours of VLA 4.8 GHz radio continuum emission (Giacani et al. 2009). The same scale is used in all maps, and the contour levels are 0.07, 0.35, 0.7, and 1.4 mJy beam$^{-1}$. The red boxes in the $+5$ km s$^{-1}$ panel represent the regions used to extract the CO lines in Figure 3. The red lines in the $+10$ km s$^{-1}$ panel indicate the locations and directions used to plot the position–velocity (PV) diagrams in Figure 4, and the two small light green circles correspond to the positions of the linear patterns in the first two panels of Figure 4.

### 3.2. Line Profiles

We choose 15 regions along the three shells (see the $+5$ km s$^{-1}$ panel of Figure 2) of G352.7−0.1 and extract the spectra from those regions. Since a radio shell marks the shock front of an SNR, any SNR–MC interaction is more likely to take place in these regions. The spectra are presented in

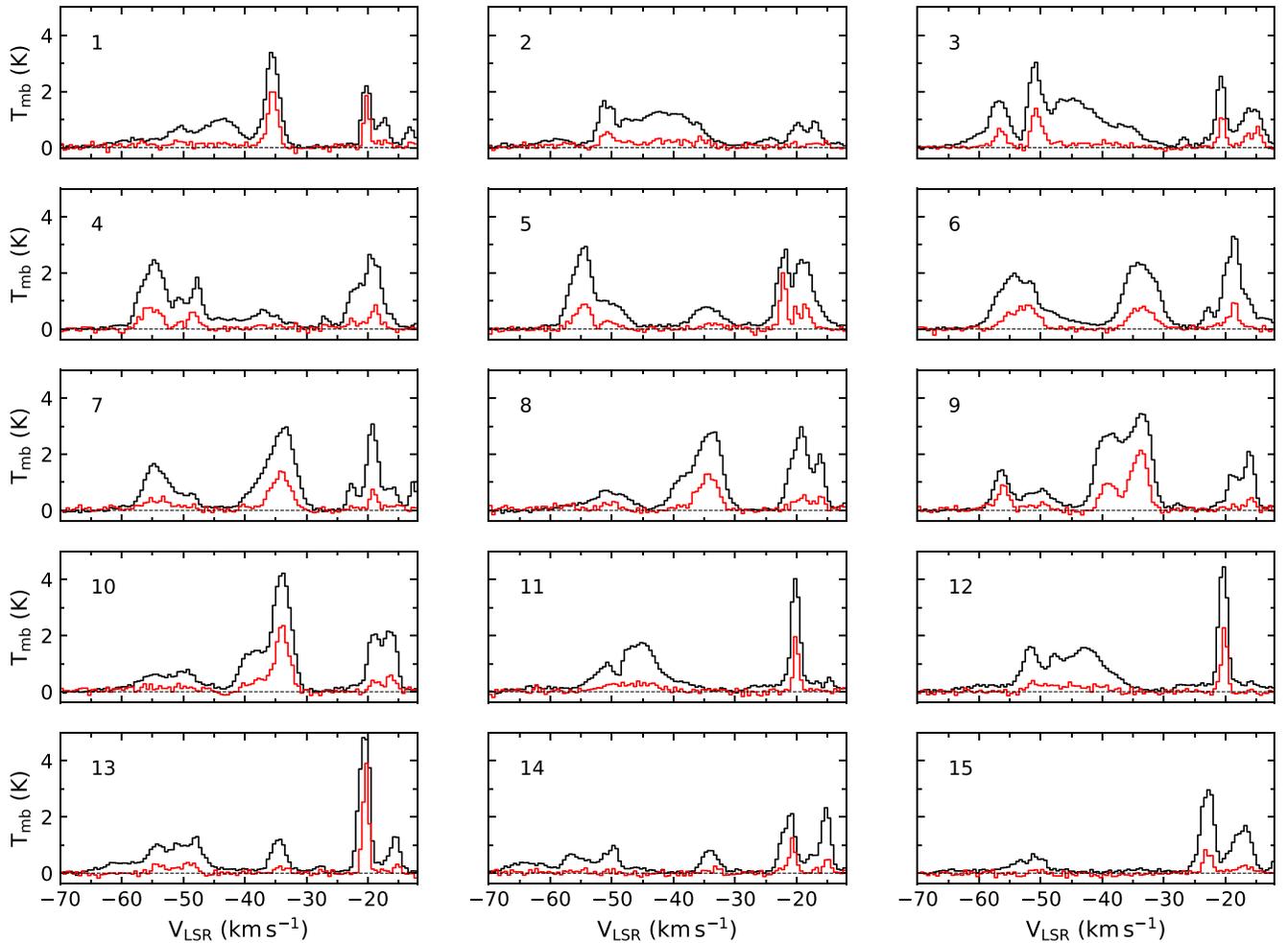

**Figure 3.** The $^{12}$CO $J = 2$–1 (black) and $^{13}$CO $J = 2$–1 (red, multiplied by a factor of two) spectra of region 1 to region 15 labeled in Figure 2 in the velocity range from $-70$ to $-12$ km s$^{-1}$. The dashed lines denote the 0 K main-beam temperature.

Figure 3. The line profiles display no significant signs of SNR–MC interaction at other velocities, so we only focus on the velocity range of $-70$ to $-12$ km s$^{-1}$.

Shocked molecular gas is characterized by broad asymmetric lines (e.g., as summarized in Jiang et al. 2010 and Chen et al. 2014). Such a signature is found in the $^{12}$CO spectra from $-50$ to $-30$ km s$^{-1}$ in region 3 (see Figure 3), while the $^{13}$CO emission is weak in this velocity interval as it is usually optically thin and much weaker than the $^{12}$CO emission in perturbed gas. In region 1, the $^{12}$CO profile seems to have a broad wing from $-60$ to $-40$ km s$^{-1}$. Despite some spatial correspondence (as mentioned in Section 3.1), other regions do not show prominent broadened $^{12}$CO lines with undetected $^{13}$CO in the same velocity interval, though there are line wings with an FWHM less than 5 km s$^{-1}$ that possibly indicate some minor turbulence.

### 3.3. PV Diagrams of the Molecular Gas

Figure 4 shows PV diagrams of $^{12}$CO across the remnant. The diagrams are made along the lines which are labeled with "1"–"6" in the 10 km s$^{-1}$ panel of Figure 2. There is a linear pattern extending from $\sim -50$ to $\sim -35$ km s$^{-1}$ at offset $\sim -1'$ in panels 1–5, which strengthens our argument that the remnant is interacting with MCs. Besides, there is an annular pattern in all six panels. Such an annular pattern in the PV diagram is often related to the expanding motion of gas.

### 3.4. Nondetection of GeV Gamma Rays

Many MMSNRs are GeV–TeV gamma-ray sources due to physical interaction with MCs (Liu et al. 2015). The first Fermi-LAT SNR catalog (Acero et al. 2016) reported a GeV candidate for G352.7$-$0.1, but it has little overlap with the radio SNR. Since it is very likely that SNR G352.7$-$0.1 interacts with adjacent MC(s), we wonder if there is gamma-ray emission from hadronic processes, in which neutral pions produced by collisions between SNR-accelerated cosmic-ray protons and dense gas decay into gamma rays.

We analyze the Fermi-LAT gamma-ray data in the energy range 0.1–500 GeV. We free the normalization and index of the catalog sources within 3° of the ROI center as well as the diffuse background components. Then we execute a fit and generate the test-statistic (TS) map. The TS value is defined as TS $= 2\log(L_1/L_0)$, where $L_1$ is the likelihood of the hypothesis being tested and $L_0$ is the likelihood of the background. All the TS values within 0°.5 from the ROI center are less than 16. So we conclude that no significant GeV gamma-ray emission is detected toward the remnant. This can give some constraints on the model parameters of the hadronic process (see Section 4.5).

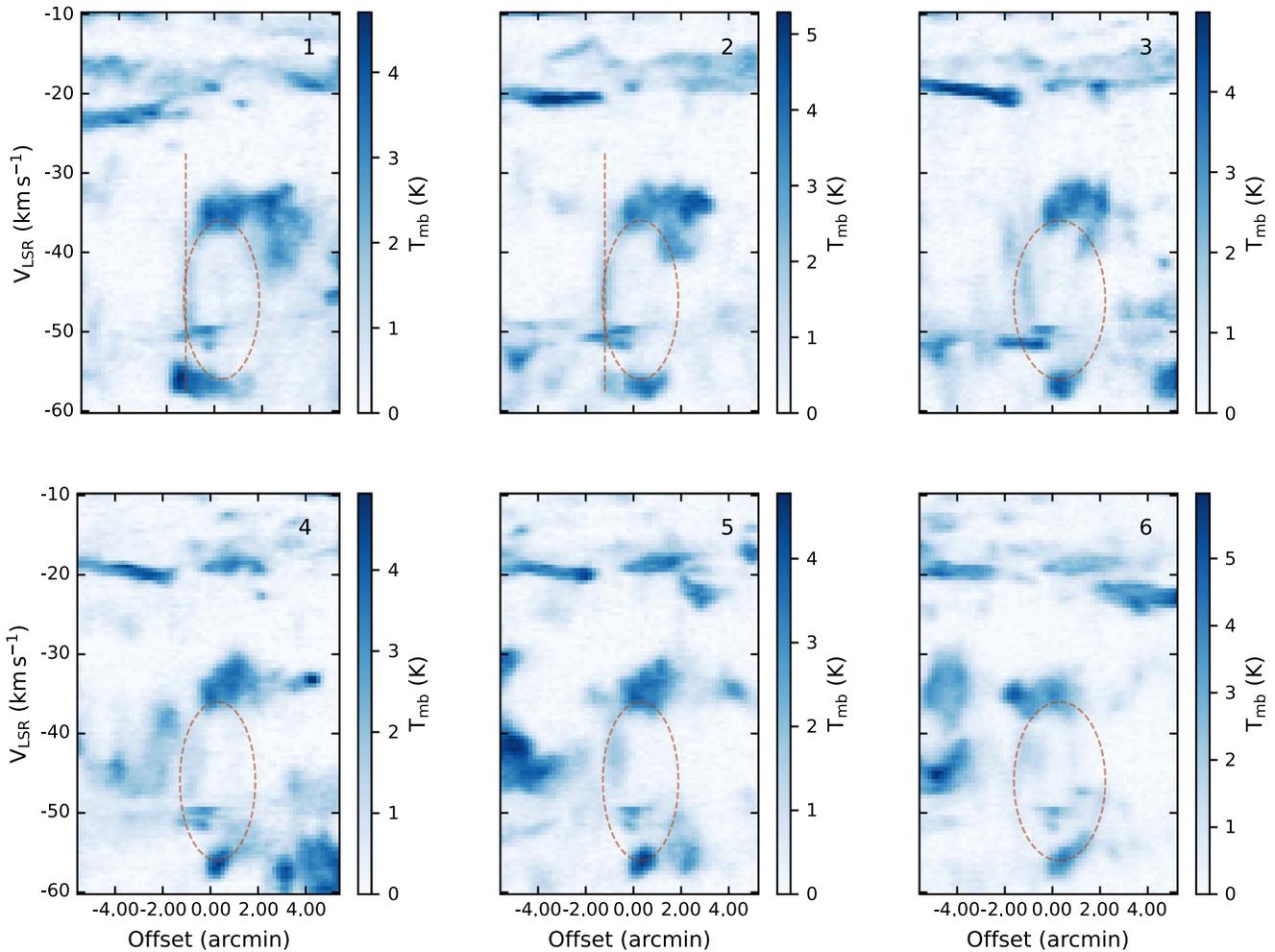

**Figure 4.** PV diagrams of $^{12}$CO $J = 2$–1 made along the lines in Figure 2. The red dashed lines in the first two panels indicate the linear pattern which corresponds to the light green circles in the +10 km s$^{-1}$ panel of Figure 2, and the ellipses in all six panels delineate the expanding structure with the short axis corresponding to the span of each black line in Figure 2 within the northeastern ring.

## 4. Discussion

### 4.1. Association of the −51 km s$^{-1}$ Cloud with G352.7−0.1

The broad wings shown in the $^{12}$CO line profiles (Section 3.2) and the PV diagrams (Section 3.3) indicate possible interactions between G352.7−0.1 and adjacent MCs. Figure 5 compares the $^{12}$CO distribution in the broad-wing interval of −48 to −37 km s$^{-1}$ with the radio emission of the remnant. Two patches are seen in the middle shell. The typical profile of the northern patch (indicated by a cyan ellipse) is a narrow line at ∼−35 km s$^{-1}$ with a blue wing (like region 1), while the typical profile of the southern patch (indicated by a white ellipse) is a narrow line at ∼−51 km s$^{-1}$ with a red wing (like region 3).

The shock is unlikely to interact with both the ∼−35 km s$^{-1}$ and ∼−51 km s$^{-1}$ clouds. The two LSR velocities correspond to two distances separated by more than 1 kpc (according to a flat Galactic rotation curve; Reid et al. 2014; Wenger et al. 2018), which is much larger than the typical scale of an SNR. Moreover, the two velocities cannot be explained by receding and approaching gas, respectively, in an expanding shell driven by a progenitor wind or the shock itself. For example, if the $\sim-51$ km s$^{-1}$ component corresponds to the approaching gas (as the velocity is more negative than $\sim-35$ km s$^{-1}$), the broadened CO emission produced by the shock behind the clouds should be a blueshifted line. However, the broadened emission at $\sim-50$ to $-30$ km s$^{-1}$ is a red wing of $\sim-51$ km s$^{-1}$.

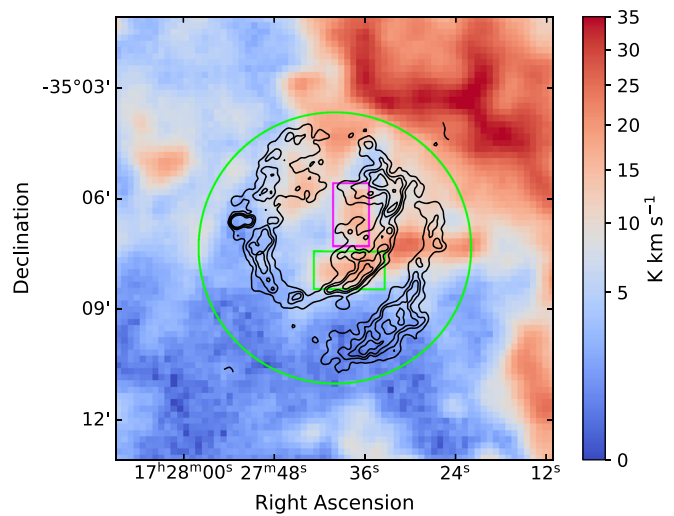

**Figure 5.** The intensity map of $^{12}$CO $J = 2$–1 integrated from −48 to −37 km s$^{-1}$. The black contour is the same as in Figure 2. The green circle indicates the region used to calculate the mass of the ambient molecular gas. The magenta and green rectangles show the northern and southern patches, respectively.

We propose that G352.7−0.1 is more likely to interact with the southern $\sim-51$ km s$^{-1}$ cloud in the south of the middle shell, because the $^{12}$CO line wing of the $\sim-51$ km s$^{-1}$ cloud (the southern patch) has a better spatial correspondence with the radio (see Figure 5) and infrared (see Pannuti et al. 2014)) morphology of G352.7−0.1 than the $\sim-35$ km s$^{-1}$ cloud (the northern patch).

We examine the CO spectral grid of the southern part of the middle shell (see Figure 9 in the Appendix). The broad line wings are obvious in the $^{12}$CO profile from $-50$ to $-30$ km s$^{-1}$ and seem to distribute along the northwest–southeast direction, just like the direction of the middle shell. Though the line profiles vary from one pixel to another, they are similar to the profile of region 3 generally.

Though the $\sim-35$ km s$^{-1}$ cloud may not interact with the remnant directly, it may be part of a wind-blown bubble associated with the progenitor (See Section 3.3), and the annular patterns in Figure 4 also favor a systemic velocity of $\sim-51$ km s$^{-1}$ rather than $\sim-35$ km s$^{-1}$. Since the $\sim-35$ km s$^{-1}$ cloud is massive, this component is more likely made up of two separate parts: (1) the gas swept up by progenitor winds and (2) the local cloud at $\sim-35$ km s$^{-1}$ along the line of sight (see Figure 10 in the Appendix for details).

### 4.2. The Distance

The distance to G352.7−0.1 is still ambiguous. Early studies applied the radio surface brightness–linear diameter ($\Sigma$–$D$) relation and put the remnant at a distance of $\geqslant 11$ kpc (Clark et al. 1973; Caswell et al. 1983; Dubner et al. 1993). The intervening hydrogen column density $N_H$ ($\sim 3 \times 10^{22}$ cm$^{-2}$) obtained from X-ray observations suggested a distance of $8.5 \pm 3.5$ kpc (Kinugasa et al. 1998). Giacani et al. (2009) derived a range between $\sim 6.8$ kpc and $\sim 8.4$ kpc based on the H I absorption profile, so they assumed the distance to be $7.5 \pm 0.5$ kpc.

According to our discussion in Section 4.1, the systemic velocity of G352.7−0.1 is $\sim-51$ km s$^{-1}$. Employing the Galactic rotation curve model with $R_0 = 8.34$ kpc and $V_0 = 240$ km s$^{-1}$ (Reid et al. 2014; Wenger et al. 2018), we derive a near kinematic distance of $\sim 6$ kpc and a far kinematic distance of $\sim 10.5$ kpc.

To determine the relative position of G352.7−0.1 with respect to the tangent point, we inspect the H I absorption profile. The regions where we extract H I source and background spectra are marked in Figure 6. The H I absorption spectrum is expressed by Tian et al. (2007) as

$$e^{-\tau} = 1 - \frac{T_{\rm off}(v) - T_{\rm on}(v)}{T_s^c - T_{\rm bg}^c}, \qquad (1)$$

where $T_{\rm on}$ and $T_{\rm off}$ are the H I brightness temperatures of the source and background regions, and $T_s^c$ and $T_{\rm bg}^c$ are the continuum brightness temperatures of the source and background regions, respectively.

Figure 7 shows the H I emission and absorption spectra of the selected regions. The most negative peak in the emission profile is at $\sim-85$ km s$^{-1}$, roughly corresponding to the tangent point (at $\sim 8.25$ kpc) velocity given by the Galactic rotation curve model. Assuming a homogeneous distribution of H I, the absorption features at $<-51$ km s$^{-1}$ (including that at the tangent point) indicate that the remnant must be located beyond the tangent point, so the far kinematic distance of $\sim 10.5$ kpc is favored. It is worth mentioning that the difference between our result and that of Giacani et al. (2009) is due to different Galactic rotation curve models and solar motion parameters (Fich et al. 1989; Reid et al. 2014). We also emphasize that H I gas is not necessarily associated with MCs. Even though they are physically associated, we might not detect an absorption signature at $\sim-51$ km s$^{-1}$, because the gases lie behind the remnant as we have discussed in Section 4.1.

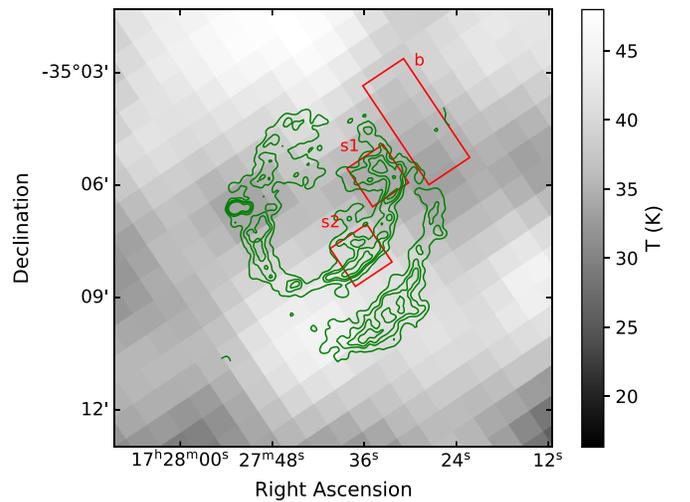

**Figure 6.** H I spatial distribution in the velocity range of $-53$ to $-48$ km s$^{-1}$, overlaid with contours of VLA 4.8 GHz radio continuum emission. The red boxes represent the source (s1 and s2) and background (b) regions used in Figure 7.

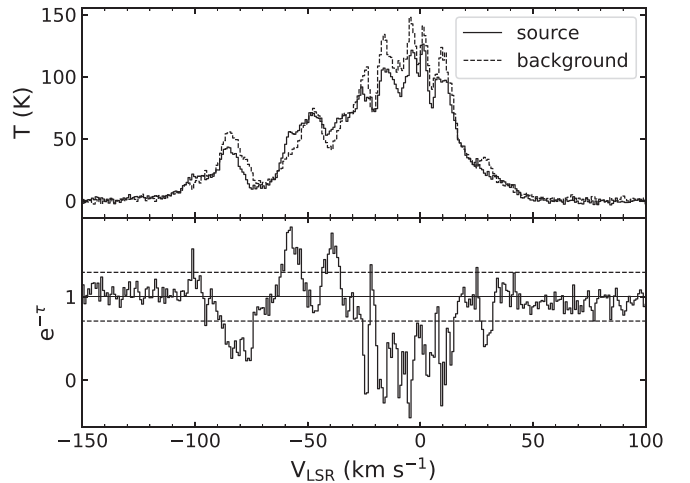

**Figure 7.** H I emission and absorption spectra of the regions marked in Figure 6. The background spectrum is extracted from region b. The source spectrum is derived by averaging the spectra of regions s1 and s2, which have the highest radio luminosities. The horizontal dashed lines represent the $\pm 3\sigma$ uncertainty level of the H I absorption spectra.

Assuming that $^{12}$CO and $^{13}$CO are optically thick and optically thin, respectively, we can derive the mass of the

$\sim -51$ km s$^{-1}$ cloud. The column density of $^{13}$CO is expressed by Mangum & Shirley (2015) as

$$N(^{13}\mathrm{CO}) \approx 3.1 \times 10^{13} \left(0.38 T_{\mathrm{ex}} + \frac{1}{3}\right) \exp\left(\frac{15.9}{T_{\mathrm{ex}}}\right)$$
$$\times \frac{\int T_{\mathrm{mb}}(^{13}\mathrm{CO}) dv (\mathrm{km\ s}^{-1})}{[48.2 - \exp(\frac{10.6}{T_{\mathrm{ex}}})] f} \mathrm{cm}^{-2}, \quad (2)$$

where $f$ is the filling factor (the fraction of the telescope beam occupied by the source), and $T_{\mathrm{ex}}$ is the excitation temperature, which can be derived from the optically thick $^{12}$CO line using $T_{\mathrm{ex}} = 11.1[\ln(1 + 11.1(T_{\mathrm{mb}}/f + 0.2)^{-1})]^{-1}$ K (Mangum & Shirley 2015). Low excitation temperatures are obtained ($\sim 3$–9 K) across the cloud if we assume the filling factor to be one, so the real filling factor should be smaller. We adopt a temperature of 10 K, which is the typical temperature of quiescent MCs. The mass of the cloud is given by $M = 7 \times 10^5 \mu m_{H_2} N(^{13}\mathrm{CO}) d^2 \Delta\Omega$, where $\mu$ is the mean atomic weight of the gas ($\mu = 1.4$), $d$ is the distance to the remnant in units of kiloparsecs, and $\Delta\Omega$ is the area of the cloud ($R \sim 220''$; see the green circle in Figure 5). Here the relation $N(H_2) \approx 7 \times 10^5 N(^{13}\mathrm{CO})$ is used (Frerking et al. 1982). Taking $d \sim 10.5$ kpc, the total mass of the $\sim -51$ km s$^{-1}$ cloud is estimated to be $\sim 1.8 \times 10^3 M_\odot$. However, we will not use this value in the following analyses. We suggest that the MCs in the velocity range of $-58$ to $-30$ km s$^{-1}$ are possibly related to the progenitor wind, so we reasonably adopt $M \sim 0.5$–$1.0 \times 10^4 M_\odot$ as the total mass associated with the remnant (see Section 4.3 for details).

### 4.3. The Double-ring Morphology and Progenitor of G352.7−0.1

Kinugasa et al. (1998) first gave an explanation for the double-ring structure of G352.7−0.1. The inner shell (the northeastern ring) originates from the SN explosion while the outer shell (the southwestern ring) is attributed to the stellar wind of the progenitor. Giacani et al. (2009) used a "barrel-shaped" scenario (Manchester 1987) to explain the special structure of G352.7−0.1, where the remnant interacts with the axially symmetric wind of the progenitor star and appears to have a biannular morphology when observed from a specific angle.

Manchester (1987) accounted for the biannular radio structure of SNRs with four possible mechanisms: (1) bipolar pre-SN winds, (2) asymmetries in the SN explosion, (3) biconical winds from a central pulsar, and (4) precessing jets from a central binary system. No compact object has been found in G352.7−0.1 so far, and there is no evidence for asymmetric SN explosion either. We then propose the possibility that G352.7−0.1 is shaped by the progenitor wind. As is shown in Figure 4, there appears to be an annular pattern in the PV diagrams of $^{12}$CO, which may imply the existence of an expanding molecular gas structure that is very likely related to the history of a wind-blown bubble. The bubble expands faster on the receding side and slower on the approaching side, suggesting an anisotropic distribution of the ambient medium. If the pre-SN wind is bipolar, it may provide an explanation for the double-ring structure of G352.7−0.1.

In addition to the possible explanations mentioned above, Toledo-Roy et al. (2014) suggested a blowout model for the remnant. In their model, the SN explodes in a spherical cloud near the cloud border. When the remnant reaches the interface between the dense cloud and the low-density ISM, a blowout is produced. Then the northeastern ring forms at the interface, while the part remaining in the cloud becomes the western shell. Their simulation agrees with the observed radio and X-ray morphology when the line joining the remnant and cloud centers is rotated 60° from the plane of the sky. However, this scenario seems unlikely as the western shell is not embedded in dense clouds according to our observations.

The wind-blown bubble also helps interpret the centrally filled thermal X-ray morphology of the remnant. When the SNR interacts with the cavity wall, the transmitted shock perturbs the MCs. Meanwhile, a reflected shock is created and reheats the interior ejecta, enhancing the thermal X-ray emission. A similar scenario has been found in SNR Kes 41 (Zhang et al. 2015).

The expanding structure in Figure 4 can help us constrain the wind parameters and infer the type of the progenitor. The mean radius and the mean velocity of the expanding structure are $R_b \sim 5.2$ pc and $V_b \sim 10$ km s$^{-1}$, respectively. The fraction of the $-35$ km s$^{-1}$ cloud that is associated with the progenitor's wind is uncertain, so we can only derive the upper and lower mass limits of the ambient gas, which are $0.5 \times 10^4 M_\odot$ (based on the velocity range of $-58$ to $-40$ km s$^{-1}$) and $1.0 \times 10^4 M_\odot$ (based on the velocity range of $-58$ to $-30$ km s$^{-1}$), respectively. The corresponding atomic hydrogen number density is $n \sim 25$–50 cm$^{-3}$. If $V_b$ derived from the PV diagram is the expansion velocity of the progenitor's wind-blown bubble, according to the bubble model by Weaver et al. (1977), the duration and mechanical luminosity of the wind are $t_w \sim 3.0 \times 10^5 d_{10.5}$ yr and $L_w \sim 2.8 \times 10^{35} (n/25$ cm$^{-3})(V_b/10$ km s$^{-1})^3 (R_b/5.2$ pc$)^2 d_{10.5}^2$ erg s$^{-1}$, respectively. If the progenitor of G352.7−0.1 is a massive star, the mechanical luminosity is typical for the main-sequence wind of an O8.5–O7 star (Chen et al. 2013). But the timescale of the expanding bubble would imply an earlier type of progenitor star. If $V_b$ represents the velocity of the transmitted shock of the SNR in the molecular gas, the SNR radius should be very similar to the bubble size prior to the impact of the blast wave on the bubble wall. Thus the radius $R_b \sim 5.2$ pc would imply a progenitor of mass $\sim 12 M_\odot$, corresponding to a B1 star (Chen et al. 2013).

If the remnant is a Type Ia SNR, asymptotic giant branch (AGB) winds can hardly explain such a high mechanical luminosity given the mass-loss rates ($\dot{M} < 10^{-4} M_\odot$ yr$^{-1}$) and velocities ($v_w < 30$ km s$^{-1}$) of AGB stars. So the existence of a progenitor wind may favor the single-degenerate model of Type Ia SNe. In this case, an optically thick wind is blown from the accreting white dwarf to stabilize mass transfer (Hachisu et al. 1996). Previous studies have found such an expanding bubble possibly produced by accretion winds in Type Ia SNRs (e.g., Tycho and N103B; Zhou et al. 2016; Sano et al. 2018). Adopting a wind velocity of $v_w \sim 1000$ km s$^{-1}$, the derived mechanical luminosity corresponds to a mass-loss rate of $\sim 10^{-6} M_\odot$ yr$^{-1}$, which is within the range given by the model of Hachisu et al. (1999a, 1999b).

Table 1
The Parameters of the Shocked Cloud in the Middle Shell at $\sim -51$ km s$^{-1}$

| $v_0$ (km s$^{-1}$) | $\int T_{mb} dv$ (K km s$^{-1}$) | $\bar{v}$ (km s$^{-1}$) | $\sigma$ (km s$^{-1}$) |
|---|---|---|---|
| −50.7 | 23.0 | −43.1 | 6.5 |
| $M$ (10 $M_\odot$) | $p_{los}$ (10$^2$ $M_\odot$ km s$^{-1}$) | $E_{k,los}$ (10$^{46}$ erg) | |
| 4.3–7.2 | 3.4–5.7 | 4.7–7.9 | |

**Note.** The first line is derived from the spectrum averaged over the cloud, and the second line is derived by summing up the value of each pixel. The mass is derived under the assumption of local thermodynamic equilibrium and optically thin shocked $^{12}$CO. We use $10^{-4}$ as the CO/H$_2$ abundance ratio, 20–60 K as the excitation temperature, and 1.4 as the molecular weight.

### 4.4. Momentum and Kinetic Energy

SN explosions inject momentum and kinetic energy into the ISM, influencing star formation and the galactic structure. Previous simulations suggest that the momentum and kinetic energy input into the ISM provided by an SN are $\sim 1-5 \times 10^5 M_\odot$ km s$^{-1}$ and $\sim 10^{51}$ erg, respectively, but there are relatively few works quantifying such feedback based on observations (Koo et al. 2020; Cosentino et al. 2022). In this section, we calculate how much momentum and kinetic energy is transferred to the molecular phase of the ISM by SNR G352.7−0.1.

The momentum and kinetic energy along the line of sight respectively read

$$p_{los} = \sum (v - v_0) \Delta M(v), \quad (3)$$

$$E_{k,los} = \sum \frac{1}{2}(v - v_0)^2 \Delta M(v), \quad (4)$$

where $v_0$ is the systemic velocity of the cloud. The total momentum and kinetic energy along the line of sight are derived by summing up the values of each velocity channel.

Equations (3) and (4) can also be written respectively as

$$p_{los} = M_{tot}(\bar{v} - v_0), \quad (5)$$

$$E_{k,los} = \frac{1}{2} M_{tot}(\bar{v} - v_0)^2 + \frac{1}{2} M_{tot} \sigma^2, \quad (6)$$

where $M_{tot}$ represents the total mass of the shocked gas. $\bar{v}$ and $\sigma^2$ are, respectively, the first and second moments of the molecular spectrum

$$\bar{v} = \frac{\sum v T_{mb} \Delta v}{\sum T_{mb} \Delta v}, \quad (7)$$

$$\sigma^2 = \frac{\sum (v - \bar{v})^2 T_{mb} \Delta v}{\sum T_{mb} \Delta v}, \quad (8)$$

where $T_{mb}$ is the main-beam temperature and $\Delta v$ is the velocity resolution of the spectra.

We fit the averaged $^{12}$CO spectra of the shocked cloud (see the green rectangle in Figure 5) and give the derived parameters in Table 1.

Assuming the momentum and kinetic energy are of the same order of magnitude for different directions, we conclude that less than 1% of the total momentum and about 0.01% of the total kinetic energy have been transferred to the molecular gas by SNR G352.7−0.1. These are very small percentages compared to older SNRs like IC 443 and W44 where $p_{sh} \sim 10^5 M_\odot$ km s$^{-1}$ and $E_{k,sh} \sim 10^{49}$ erg (Koo et al. 2020).

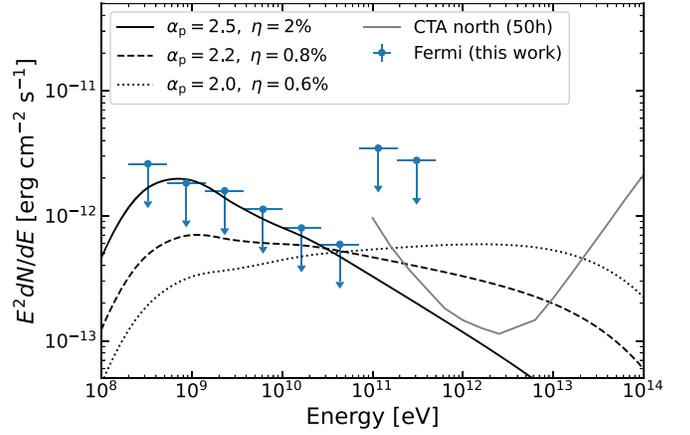

**Figure 8.** The fitted SED of G352.7−0.1. The 50 hr sensitivity of CTA (Cherenkov Telescope Array Consortium et al. 2019) is also plotted.

The age of G352.7−0.1 will be discussed in another paper (L.-X. Dang et al. 2023, in preparation).

### 4.5. The Upper Limit of the Gamma-Ray Flux

To evaluate the upper limit of the gamma-ray flux of G352.7−0.1, we add a source described by a power-law spectrum with a spectral index of 2.0 to the model. Then we use the sed method in Fermipy to derive the 99% confidence level upper limit of the energy flux for each bin.

We assume that the SNR shock converts a fraction $\eta$ of the explosion energy $E_{SN}$ into relativistic protons. The shock-accelerated protons are assumed to have a power-law spectrum with a high-energy cutoff of

$$dN/dE = A_p (E/E_0)^{-\alpha_p} \exp(-E/E_c), \quad (9)$$

where $\alpha_p$ and $E_c$ are the power-law index and high-energy cutoff, respectively. The normalization $A_p$ is determined by the total energy $\eta E_{SN}$ in the protons with energies above 1 GeV. For the explosion energy, the canonical value of $E_{SN} = 10^{51}$ erg is adopted. Due to the lack of constraints of the upper limits, the cutoff energy is fixed as $E_c = 1$ PeV and three proton indices of $\alpha = 2.0, 2.2$ and $2.5$ are considered. Cosmic-ray particles can escape into the ambient clouds that are not interacting with the shock directly, so all the molecular gas in the vicinity of the remnant can be the target gas for proton–proton hadronic interactions. Therefore, we take $n_t \sim 50$ cm$^{-3}$ as the atomic hydrogen density of the target gas being collided by the energetic protons, which is the upper limit of the ambient gas density (see Section 4.3). Then, we use the python package *naima*[6] (Zabalza 2015) with the proton–proton cross section from Kafexhiu et al. (2014) to calculate the hadronic gamma rays. In order to avoid exceeding the GeV upper limits, we find that the energy conversion fraction should be less than 0.6%, 0.8%, and 2%$(n_t/50$ cm$^{-3})^{-1}(d/10.5$ kpc$)^2$ for $\alpha_p = 2.0, 2.2,$ and $2.5$, respectively. The fitted SED is displayed in Figure 8 in which the 50 hr sensitivity of the Cherenkov Telescope Array (CTA; Cherenkov Telescope Array Consortium et al. 2019) is also plotted. As can be seen, this source may be detected by CTA in the TeV domain if the power-law index of the accelerated protons is less than 2.5. Thus one can expect that the CTA observations in the future will give stronger constraints on the hadronic process.

---
[6] https://naima.readthedocs.io/en/latest/mcmc.html

## 5. Summary

We investigate the molecular environment of SNR G352.7−0.1 by performing $^{12}$CO $J = 2$–1 and $^{13}$CO $J = 2$–1 observations with APEX. Then we analyze Fermi-LAT data to search for gamma-ray emission from the remnant. The main results are as follows.

1. The broadened $^{12}$CO profile found to the south of the middle shell provides evidence of interaction between G352.7−0.1 and MCs at ∼−51 km s$^{-1}$, which strengthens the link between thermal composite SNRs and SNR–MC interactions.
2. No gamma-ray emission is detected in the energy range 0.1–500 GeV.
3. The distance to the remnant is about 10.5 kpc based on the systemic velocity of ∼−51 km s$^{-1}$.
4. The momentum and kinetic energy of the shocked molecular gas along the line of sight are ∼$10^2 M_\odot$ km s$^{-1}$ and ∼$10^{46}$ erg, respectively.
5. We have found signatures of expanding motion of the molecular gas around G352.7−0.1, which is possibly related to the progenitor wind.

## Acknowledgments

This publication is based on data acquired with the Atacama Pathfinder Experiment (APEX) under program ID 0103.D-0387 (A). APEX is a collaboration between the Max-Planck-Institut fur Radioastronomie, the European Southern Observatory, and the Onsala Space Observatory. We thank G. Dubner for providing us with the 4.8 GHz VLA image of G352.7-0.1. We acknowledge the support from National Key R&D Program of China under Nos. 2018YFA0404204 and 2017YFA0402600, and NSFC grants under Nos. 12273010, 12173018, 12121003, and U1931204. P.Z. acknowledges the support from Nederlandse Organisatie voor Wetenschappelijk Onderzoek (NWO) Veni Fellowship, grant No. 639.041.647.

*Software*: astropy (Astropy Collaboration et al. 2013, 2018), GILDAS (Pety 2005; Gildas Team 2013), DS9 (Smithsonian Astrophysical Observatory 2000; Joye & Mandel 2003), and Fermipy (Wood et al. 2017).

## Appendix A
## Grid of the CO Spectra

Figure 9 shows the grid of $^{12}$CO and $^{13}$CO spectra for the southern part of the middle shell. The broad line wings which suggest SNR–MC interaction are obvious in the $^{12}$CO profile from −50 to −30 km s$^{-1}$ and seem to distribute along the northwest-southeast, just like the direction of the middle shell.

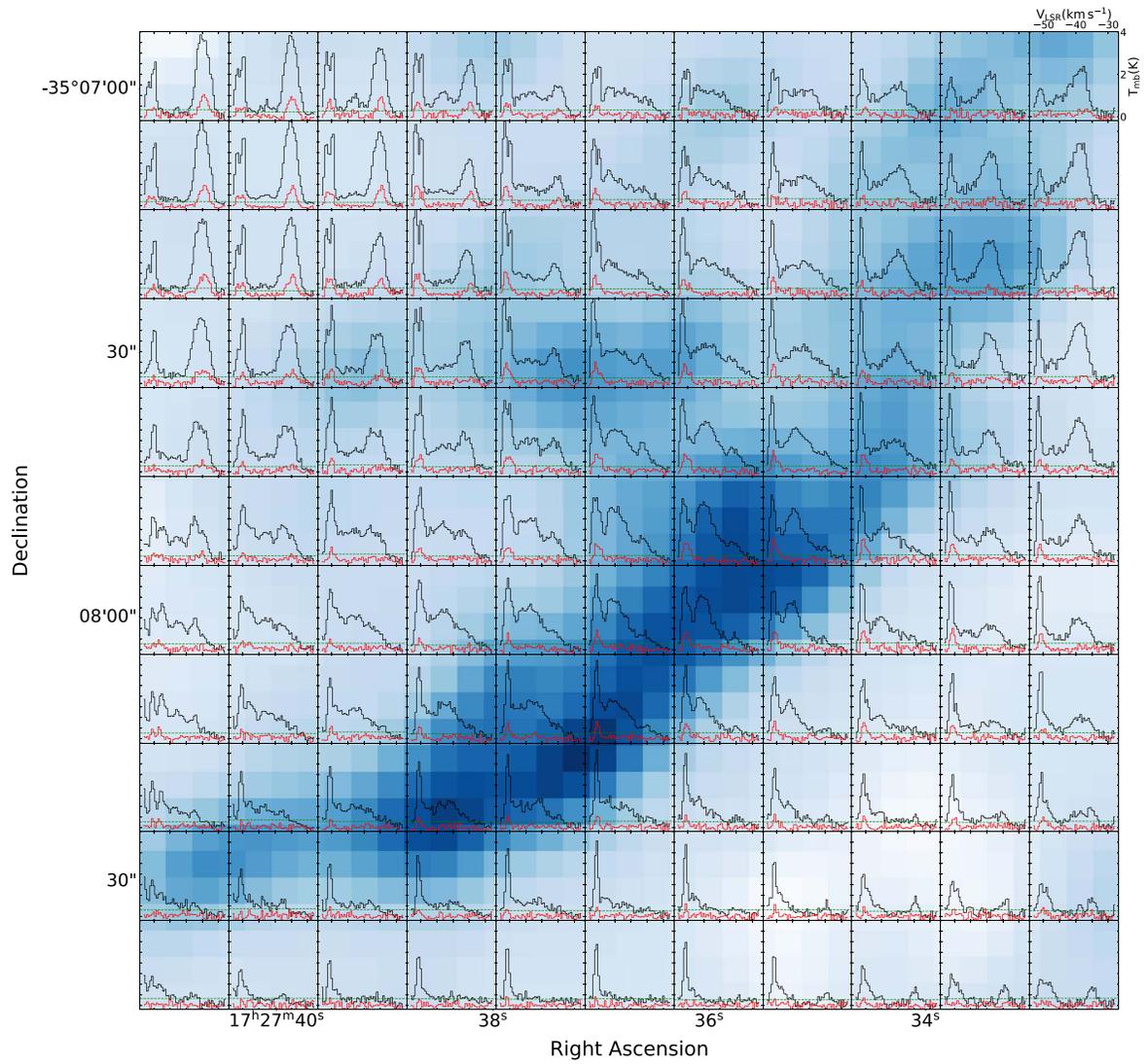

**Figure 9.** Grid of $^{12}$CO (black) and $^{13}$CO (red) spectra for the southern part of the middle shell in the velocity range from $-53$ to $-28$ km s$^{-1}$. The dashed horizontal lines denote the 3$\sigma$ level of $^{13}$CO. The background is the VLA 4.8 GHz image.

# Appendix B
## $^{12}$CO Moment Maps

Figure 10 shows the zeroth and the first moment maps of $^{12}$CO in the velocity range of −40 to −30 km s$^{-1}$. A velocity shift can be seen across the $^{12}$CO gas in the first moment map, which indicates that the −35 km s$^{-1}$ cloud may consist of more than one component.

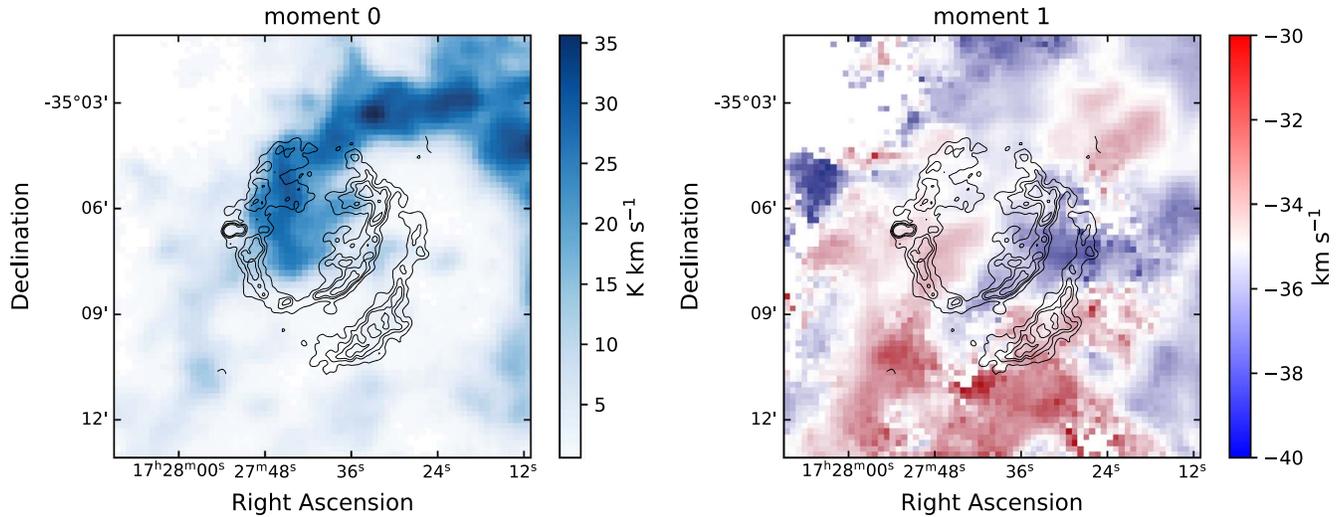

**Figure 10.** The zeroth (left panel) and the first (right panel) moment maps of $^{12}$CO in the velocity range of −40 to −30 km s$^{-1}$, overlaid with contours of VLA 4.8 GHz radio continuum. A velocity shift can be seen across the $^{12}$CO gas, which indicates a complex composition of the ∼−35 km s$^{-1}$ cloud in a large field not limited in the extent of the SNR.

## ORCID iDs

Qian-Qian Zhang https://orcid.org/0000-0003-0853-1108
Ping Zhou https://orcid.org/0000-0002-5683-822X
Yang Chen https://orcid.org/0000-0002-4753-2798
Xiao Zhang https://orcid.org/0000-0002-9392-547X
Wen-Juan Zhong https://orcid.org/0000-0003-3717-2861
Xin Zhou https://orcid.org/0000-0003-2418-3350
Zhi-Yu Zhang https://orcid.org/0000-0002-7299-2876
Jacco Vink https://orcid.org/0000-0002-4708-4219